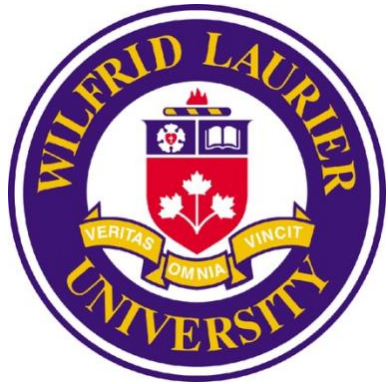

# EC 681: RESEARCH PAPER AND SEMINAR

## Final Paper

STUDENT: ASMAR ALIYEVA
SUPERVISOR: DR. MICHAEL BROLLEY
COURSE INSTRUCTOR: DR. RANDALL WIGLE

# Insider Ownership and Dividend Payout Policy:

# The Role of Business Cycle

# Table of Contents

## Abstract

We investigate how the relationship between managerial stock incentives and the dividend payout policy is impacted by the business cycle by using the data of S&P 1500 companies during 2000-2018. We find a strong negative relationship between managerial stock options and annual dividend payouts of companies for the full sample. Although the direction of the relationship is also negative for the recession period, the coefficient is found to be insignificant. We also find that the mentioned relationship may vary during the recession depending on the size of the company. The impact of stock options on the dividend payout is negative for medium-sized companies and the coefficient is both economically and statistically significant. The direction of impact changes for large-cap companies indicating to deterioration of the CEO voting power in those companies and less agency problem. We also determine that the percentage of shares held by the CEO has a positive impact on annual dividends distributed for large-cap companies, whereas this relationship changes in times of recession.

## 1. Introduction

The impact of managerial stock incentives on the dividend payout decision of companies depends on individual components of the compensation package. It has been suggested by the literature that stock options contribute negatively to the dividend distribution, whereas the impact of shares held by management and other forms of compensation were found to be ambiguous or insignificant (Fenn and Liang, 2001). However, the literature does not explain how this relationship changes in times of economic contraction and expansion and their impact on managerial incentives. What is the role of the business cycle in how managerial stock incentives impact the dividend payout decision of companies?

In times of economic expansion, when shares prices are performing well, the probability of options to be out-of-the-money is low. Consequently, management who holds stock options does not have a stimulus to undertake risky investments. In contrast, economic contraction is associated with large volatility in stock prices and returns. The high sensitivity of the value of the managerial compensation portfolio, especially stock options, to stock price and return incentivize managers to undertake risky investments and distribute less dividends to improve the chances of their options to be in-the-money. This may not always serve shareholders' best interests (Coles et al, 2006).

The signaling power of dividends is found to be stronger in times of economic contraction as investors value dividend-paying companies more in a declining market environment. Moreover, dividend-paying companies are found to outperform non-dividend paying companies at all stages of the business cycle and this difference is even significant in a downturn period (Fuller and Goldstein, 2005). Although the economic contraction period is associated with excessive positive sentiment of investors to dividend payouts in comparison with the period of advancing markets, the management may choose to decrease dividend payments, thereby freeing up cash to finance risky investments. Thus, it is crucial to investigate how the business cycle impacts managerial stock incentives and agency problems in the company.

To analyze the role of the business cycle in how managerial stock incentives impact the dividend payout policy of the company, we use the data of S&P 1500 companies. We collect the data regarding the number of shares and exercisable stock options held by the CEO, which are the variables that measure insider ownership, from *Execucomp* database. Next, we collect the data regarding the dependent variable, an annual dividend cash distribution of the company, from CRSP database. We exclude financial, telecommunication and utility companies, and companies with special/residual dividend distribution from the sample. We also include other firm-specific determinants of dividend policy such as profitability, investment, leverage and the growth rate of the company. The total final number of companies used in the analysis is 1038. The date of observations ranges from 2000-2018. Next, we incorporate the business cycle variable in my analysis by a using dummy variable to be equal to 1 if the recession was experienced in that year and 0 otherwise.

The hypothesis of this paper is to determine whether stock options negatively impact the annual dividend payout of companies during a recession. As for methodology, we focus on the OLS regression model with year and industry fixed effects. The advantage of this method is that it eliminates time-invariant year and industry effects and presents more robust results. The results of this paper will be useful for institutional investors who prefer dividend distribution, especially during contraction, and will help them to determine a more appropriate investment portfolio.

There are two main findings of this paper. First, the findings of this paper support the existing literature by determining a negative impact of stock options held by a CEO on a percentage change of annual dividend distribution for the full sample period. Although this impact is still negative for the recession period, the coefficient was found to be statistically insignificant. A possible reason behind the insignificance of the coefficient is the varying degree of the voting power of CEO depending on company size, which leads to the second finding of the paper. We find evidence supporting the change in the direction of the impact of options held by CEO on dividend distribution during the recession period depending on company size. This relationship is significantly negative for medium-sized companies and positive for large-sized companies. This indicates to possible loss of CEO voting power over the dividend distribution decision of the company.

The paper is organized as follows: Section 2 discusses existing relevant theory and research papers in the literature, Section 3 outlines the econometric model and methodology used for the analysis, Section 4 presents data description and variables used, Section 5 highlights the main findings of the paper and Section 6 concludes.

## 2. Literature Review

This research will contribute to three areas of literature: managerial stock incentives and dividend policy, managerial stock incentives and risk-taking, dividend policy and factors impacting it. Related researches about the mentioned areas of literature are presented and discussed below.

### 2.1. Managerial stock incentives and dividend policy

Bhattacharyya (2007) develops a model that suggests that insider ownership is expected to be negatively correlated with dividend payout policy of companies. According to the model agents (i.e. management) are divided into high-ability and low-ability types such that high-ability agent has access to profitable investments that company can invest in, whereas, a low-ability agent does not have enough skill to have access to those investment opportunities. The model assumes that principals (i.e. shareholders) know the type of the agent and construct such a compensation portfolio that high-ability agent gets high managerial stock incentives and low-ability agents get low managerial stock incentives. As a result, high-ability agents who possess a high degree of insider ownership invest available excess cash to profitable projects and distribute less dividend to principals, whereas, low-ability agents with low insider ownership choose to distribute excess cash in the form of dividends as they lack the access to profitable investments.

Bhattacharyya et al (2008) found empirical evidence that supports Bhattacharyya (2007) by applying the model to US data. They use the earnings retention rate (which equals 1 minus dividend payout rate) as a dependent variable and regress it on various components of managerial compensation, dividend distributed and retained earnings of the company. The results of the Tobit regression conclude a statistically significant positive impact of insider ownership on the retention rate. This implies that companies with high insider ownership experience a high earnings retention rate and low dividend distribution to shareholders. The results were unchanged even after adding control variables such as capital expenditures of company, market to book value ratio and debt to equity value ratio. The same analysis and model were applied to publicly traded companies in New Zealand (Anderson et al, 2018). The results of the paper were consistent with the theory developed by Bhattacharyya (2007) and the empirical results of Bhattacharyya et al (2008).

Fenn and Liang (2001) examine a similar relationship by using data of 1100 companies included in the S&P composite index during 1993 - 1997. The authors used four separate Tobit regressions to investigate the direct impact of shares and stock options held by management on dividend policy of the company and used firm-specific control variables such as firm size, free cash flow, market-to-book value ratio, and leverage ratio. The authors also establish a negative relationship between stock options held in compensation portfolio and dividend distributions. This result also concurs with the findings of Lambert et al (1989). However, the impact of shares and other bonuses held by management on dividend payout is found to be insignificant.

Although the paper determines negative relationship among variables of interest, it does not consider the impact of business cycle and company size.

The recent study by Geiler et al (2016) investigated the impact of managerial compensation of dividend payout policy by using the data of UK based companies listed on the London Stock Exchange during 1996-2007. They used data for components of CEO remuneration and compensation package, tax rate, market sentiment, and company-specific control variables to investigate their impact on dividend policy. The results of quantile regression, multinomial logit model and probit regressions indicate the significant negative impact of stock options held by a CEO on dividend distribution in all model regression outputs. Meanwhile, salary and bonuses are found to have positive relationships with dividend distribution. The authors also accounted for possible endogeneity problems in the model. Profitable companies may have enough ability and excess cash to pay dividends and increase managerial compensation. In this case, it is not managerial compensation that determines dividend payouts, but, rather, company profit that determines both variables. If this is true, the coefficients predicted by the model are biased and do not reflect the true relationship between variables. The authors use industry adjusted mean of each component of compensation portfolio as an instrumental variable for managerial compensation and regress it on past stock returns of the company, its ROA and industry adjusted return and added predicted values into the main regression model. The results did not change even after using Instrumental Variables, Two-Stage Least Square model and multiple robustness checks. Stock options held by management are expected to be the main reason for the "disappearing dividends" phenomenon suggested by Fama and French (2001). Management deliberately decreases dividend payments to protect the value of stock options and equity-based compensation.

The study conducted by Brown et al (2007) investigates the influence of the change in personal tax rate imposed on dividend income, specifically the 2003 Tax Cut policy, on dividend policy and use the data of 1,700 publicly traded US companies. Companies with higher insider ownership increased dividend distribution, in response to the 2003 Tax Cut, more than companies with low insider ownership, keeping other determinant variables constant. The authors reveal an increase in agency problem after the 2003 Tax Cut policy, as companies, where the majority of shareholders preferred capital gains over dividend payouts, are also experienced a dividend increase.

In contrast to the above-discussed literature, research conducted by White (1996) finds evidence of a positive impact of insider ownership on dividend payout policy. The author used data of 62 companies from oil/gas, food processing, and defense/aerospace industries (where companies experience positive excess cash flow to firm and expected agency problems are high) and investigates the impact of dividend provisions in compensation package on the dividend distribution decision of management. The result of the logistic regression model reveals a positive and significant relation between dividend provisions and actual dividends paid.

The existing literature about the impact of managerial compensation on dividend policy is the following: if stock options constitute a large fraction of compensation package, the impact of the managerial stock incentives on dividend policy is expected to be negative, whereas, the impact of salary and other bonuses are expected to be either positive or insignificant. The goal of this research paper is to extend existing literature and investigate the role of the business cycle in the relationship between insider ownership and dividend policy. The underlying motivation to investigate this question is the role of stock incentives in managerial appetite for risk-taking.

**2.2. Managerial stock incentives and risk-taking**

Managerial compensation is usually constructed such that the value of the compensation package fundamentally depends on company stock performance and return. Coles et al (2006) conclude that the more sensitive the value of the managerial stock compensation to stock volatility, the higher the possibility of risky policy choice by management. The authors analyze the US publicly traded companies during 1992-2002 by selecting a percentage change in the dollar value of managerial wealth for every 0.01 change in the stock price of the company returns as a sensitivity variable; and research and development expenditure, capital expenditure, segmentation of company earnings and degree of leverage as a proxy for firm risk. The authors found that the higher the sensitivity factor, the more management implements risky policies by investing more into R&D, employing a high degree of leverage and spending less into property plant and equipment. This indicates to increase in risk-taking.

The existence of stock options in the managerial compensation package is determined to be the major reason for risk-taking. The fundamental structure of an option is such that the value of it can be either zero (when the stock price drops below a specified threshold) or some positive amount (which is equal to the difference between stock price and conversion price determined by the contract). As the associated loss with options is limited (which is zero), management will undertake risky investments and policy choices to keep the price of stock high, which implicitly leads to high option value. Options incentivize management to use share repurchases, which leads to an increase in stock price and option value, in an excessive and improper manner. In contrast to stock options, the potential loss in share value is not limited. Thus, a larger portion of stock shares in compensation portfolio will also result in the expansion of firm risk by management (Wright et al, 2007).

The literature suggests that managerial stock incentives, especially, stock options are associated with large risky investments and policy choices. This is also true at large fractions of stock shares held by management. Although downside loss of risky investments is limited under good economic conditions, in times of economic downturn, when management wants to keep the share price high, those investments may be realized at the expense of dividend distribution to shareholders. Thus, the impact of managerial stock incentives on dividend policy is expected to change depending on the stages of the business cycle.

**2.3. Dividends and factors impacting it**

The paper will also contribute to broad literature about dividends and factors impacting it. Thus, the results of the paper may be especially valuable for investors appreciating dividend payments as the dividend is one of the two channels through which shareholders get benefits from their investment to the company. It is especially preferred by institutional investors that individual investors, because of associated get tax benefits (Miller and Scholes, 1982). The fact that institutional investors invest more in dividend-paying companies than non-dividend paying ones creates a "dividend clientele" effect for the company. Moreover, dividend-paying companies are perceived to be better performers and financially healthy which often decreases the need for monitoring by investors (Allen et al, 2000).

Fama and French (2001) analyze company-specific characteristics and dividend payout decisions of publicly traded companies during 1926 – 1999. Authors use a logit regression model to determine the impact of profitability, growth rate of assets and market – to book value ratio on the probability of dividend payout decision of companies. Companies with large company size and profitability are found to pay more dividends, whereas, companies with large investment opportunities pay less dividends. It is also expected that mature firms will pay more dividends that companies that are in the growth stage of their life cycle. However, the declining trend in dividend distributions has been detected starting from 1978 which indicates to lower propensity to pay among company managers. Even companies with high profitability and company size are found to decrease dividend distributions or not to pay at all. Even share repurchases that became a substitute for dividend distributions cannot explain this large declining trend in dividend payment.

Economic theory suggests a few possible reasons behind declining dividend payments. First, dividend distribution means low retained cash within the company which may force management to issue debt and uses external financing in the future. The second possible problem of dividend payout for management is its sustainability as once the dividend is issued, it requires to be maintained. Meanwhile, it is the existence of stock options held by management that may account for a large decrease in dividend distribution as it has been determined that the value of equity, especially, stock options is negatively correlated with dividend payouts (Lambert et al, 1989).

The theory developed by Miller and Modigliani (1958) assumes that shareholders are indifferent between the dividend distributions, capital gains, and share repurchase. However, the study conducted by Fuller and Goldstein (2005) determined that shareholders pay large attention to a dividend policy of firms and that dividends matter more in the period of economic downturn. The authors used the data for stock returns of S&P 500 companies during 1970-2000 and detected that companies who pay dividends significantly outperformed non-dividend paying companies. They used a difference in difference model by taking the difference in stock returns of dividend-paying and non-dividend paying companies as a dependent variable and regressed it on various models that control for risk (such as CAPM, Fama and French three-factor model, and Fama-MacBeth model) under two economic conditions: economic turbulence and economic stability. Even after controlling for firm-specific factors such as size, cash flow to the firm, the authors found a significant difference in stock returns of the two groups of companies in times of economic decline. This is evidence that investors value dividend-paying companies more in times of economic turbulence.

There are three economic explanations for this phenomenon. An increase in dividends has more signaling power during periods of a downturn as it indicates financial strength and belief of management into the future prosperity of the company. Second, it may signal that management does not invest in non-profitable projects and, rather, chooses to distribute cash to shareholders. Third, the distribution of dividends in times of financial turbulence contributes to the reduction of agency problems by decreasing the amount of excess cash. Consequently, management is expected to continue to pay dividends in times of downturn and take advantage of a positive shareholder attitude.

## 3. Empirical Model and Methodology

To investigate the impact of the business cycle on the relationship between managerial stock incentives and dividend payout of the firm, we use the following regression model:

*Dividend Change* = $\beta_0$ + $\beta_1$ *(Shares)* + $\beta_2$ *(Options)* + $\beta_3$ *(RecShares)* + $\beta_4$ *(RecOptions)* +

$\beta_5$ *(TCompensation)* + $\beta_6$ *(Recession)* + *Control Variables*

where:

*Dividend Change*: Annual Percentage Change in Dividend Distribution; *Shares*: Number of Shares held by the CEO as a Percentage of Shares Outstanding; *Options*: Number of Options held by the CEO as a Percentage of Shares Outstanding; *RecShares* and *RecOptions*: Interaction Term of Shares and Options Variables with Recession; *TCompensation*: Log of Total Annual CEO Compensation; *Recession*: Dummy variable equals 1 if the Recession was experienced in that year; 0 otherwise; *Control Variables*: Firm-Specific Leverage, Profitability, Investment, and Growth variables.

WEuse 3 estimation models in my analysis: OLS regression, OLS regression with year and company fixed effects and OLS regression with year and industry fixed effects. The regression model with company fixed effects controls for systematic differences among companies and firm characteristics that are unchanged over time. However, this model deteriorates the estimation of variation of the variable of interest on a firm level. To mitigate this issue and eliminate selection bias, we focus on the regression model with industry fixed effects, which is widely used in the literature (Fenn&Liang, 2001). Thus, my model assumes no company-specific heterogeneity within the industry.

## 4. Data Description

### 4.1. Sample Collection

To analyze the impact of the business cycle on the relationship between managerial stock incentives and dividend payout decision, we use the data of 1500 companies listed on S&P 1500. This includes S&P 400 Mid-Cap, S&P 600 Small-Cap and S&P 500 Large-Cap companies which constitute almost 90% of US stock market capitalization and is a good proxy for analysis of overall market dynamics. We have excluded financial, utilities, telecommunication companies, and companies with special/residual dividend policy where Standard Industrial Classification (SIC) codes were retrieved from the database of Securities and Exchange

Commission (SEC). The underlying reason behind excluding financial firms is a high degree of leverage and uncertainty of dividend payout (Fama&French, 1992). Utility and telecommunication firms are largely controlled by the government and their operations serve to improve social welfare rather than increase company revenue. The third type of companies that were excluded are companies with special/residual dividend payout as volatility in their payout exists by design rather than by a choice and decision of management. The final number of companies used for analysis is 1038.

The time period that we use in the analysis ranges from 2000 till 2018 and covers the period of the Dotcom Bubble and the Great Recession. The reason for using the mentioned period is the availability of data and the significance of missing values in the years before 2000. The total number of annual observations used in the analysis is estimated to be 19,722. We obtain business cycle duration data from the National Bureau of Economic Research (NBER). NBER reports 2001, 2008 and 2009 as the years of business contraction.

**4.2. Dividend payout**

WEcollect the data regarding dividend payout from the CRSP database. As mentioned above, special/residual dividend payouts were excluded from the sample and the number of excluded observations is not significant. This indicates that the number of companies that apply residual dividend policy decreased over time and this finding is consistent with DeAngelo et al. (2000). Companies may have different dividend distribution policies based on quarterly, semi-annual and annual distributions. Those dividends were annualized to keep consistency and comparability among observations and companies.

The measure of dividend payout variable in the model is an annual percentage change of dividend distributed per share outstanding of the company. To compute this variable, we apply the following formula:

$$Average\ Annual\ Dividend\ Change = \frac{Dividend_t - Dividend_{t-1}}{Dividend_{t-1}} * 100,$$

where *Dividend* is the annual cash dividend distributed per share outstanding.

The lowest average annual dividend change was observed in years of economic contraction (Figure 1) which indicates that the majority of the companies in the sample tend to either decrease dividend payout or keep it unchanged in times of an economic contraction. The opposite is true for the period of economic expansion when the relative change is high. The maximum percentage increase was reached in 2012 which indicates to the recovery of the market from recession.

**Figure 1: Average Annual Percentage Change in Dividend Payout**

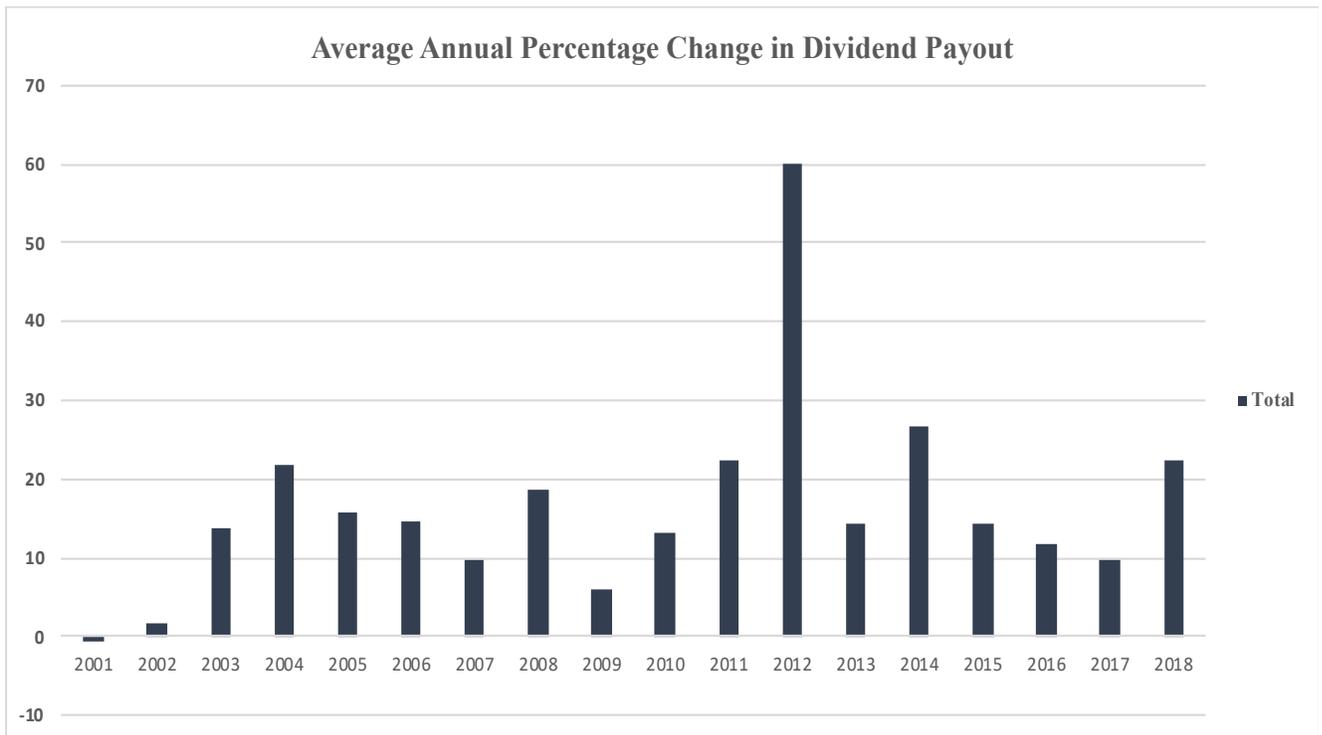

### 4.3. Insider ownership and managerial stock incentive variables

The variables that measure the degree of insider ownership is the number of shares and stock options held by management divided by the total number of shares outstanding. We collected the data from the EXECUCOMP database, and the data is on an annual basis. We used CEO data as a measure of managerial ownership which is a generally used proxy in the literature (e.g. Berger et al [1997], Mehran et al [1998]). The third variable used to measure the impact of managerial compensation on dividend policy is the logarithm of total annual compensation paid to CEO which includes salary, bonus, and other additional payment.

As Table 1 shows, on average CEO holds 2.5% of total shares outstanding, whereas, this number for options is 0.25%. The reason why the percentage of shares is almost 10 times higher than the percentage of options owned is they include only exercisable options. A similar ratio was found by Berger et al (1997) where the percentage of options held by CEO was 0.17% and 15 times less than the percentage of shares held (2.7%).

**Table 1: Descriptive Statistics**

| Variable | Obs | Mean | Std.Dev. | Min | Max |
|---|---|---|---|---|---|
| Shares | 14862 | 2.499 | 6.903 | 0 | 77.56 |
| Options | 8968 | .243 | .484 | 0 | 18.206 |
| TCompens | 15682 | 6.577 | .525 | 0 | 8.778 |
| Growth | 17654 | 9.262 | .755 | 4.663 | 11.902 |
| Investment | 17574 | 5.186 | 6.267 | -3.27 | 298.956 |
| Leverage | 17107 | .499 | .725 | .001 | 74.934 |
| Profitab | 17638 | 9.947 | 61.053 | -7956.845 | 92.041 |

Managerial Shares are highly positively skewed (Figure 2, a). CEO holding of Shares is between 0-5% of shares outstanding for more than 80% of the sample (Figure 2, b). 5% of the sample holds between 5-10% of shares outstanding and holdings of the remaining 15% are distributed among different percentages. A similar distribution is true for the number of options held by the CEO to the total number of shares outstanding (Figure 3, a). The number of options held by 85% of the sample is between 0-0.5% of shares outstanding (Figure 3, b) and the remaining part of the sample is evenly distributed. A similar distribution was reported by Fenn&Liang (2001). The number of observations of the percentage of shares held by the CEO is significantly more than the number of observations for options. This is because the inclusion of shares in the managerial compensation is more frequent and usual rather than the inclusion of options.

**Figure 2: Distribution of Managerial Shares**

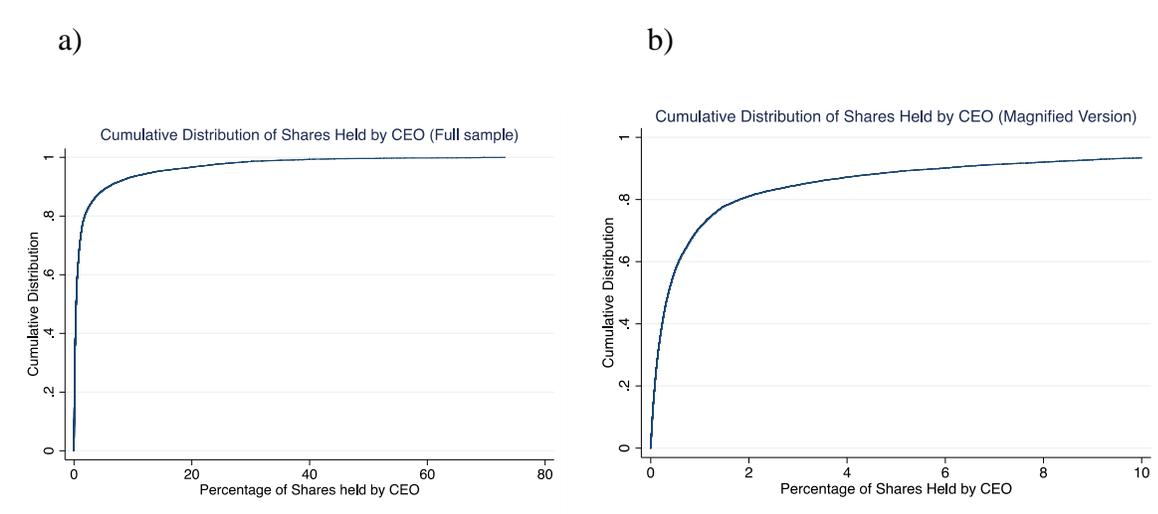

a)  b)

**Figure 3: Distribution of Managerial Options**

a)                                            b)

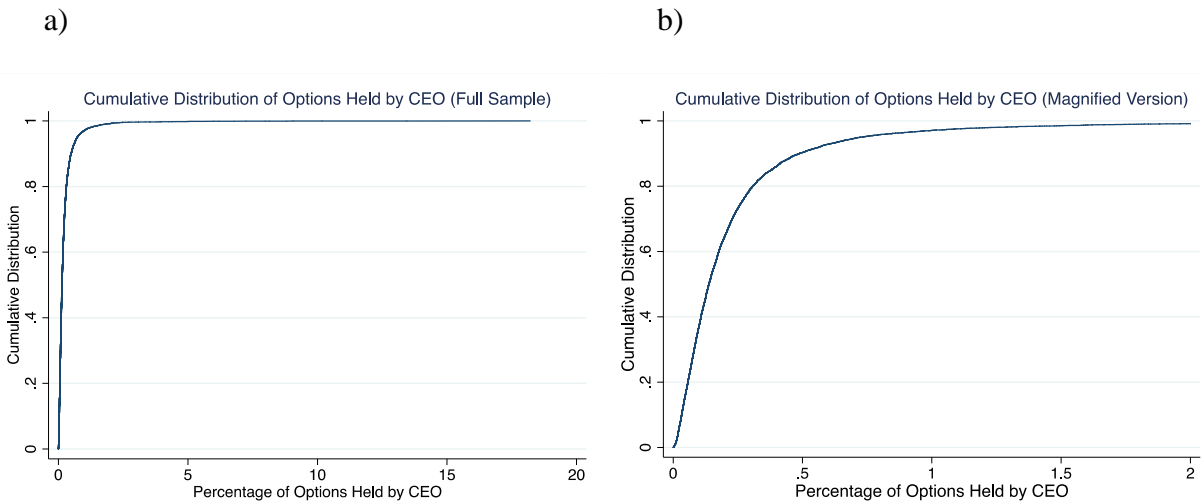

## 4.4. Firms Specific Control Variables

The dividend payout decision also depends on many other firm-specific variables. We use net free cash flow as a percentage of total assets to control for the impact of company earnings on dividend distribution. Jensen (1986) shows that companies with high positive earnings are expected to pay dividends to decrease the probability of investment in projects with non-positive NPV and to decrease agency cost. Moreover, we use the capital expenditure of the company expressed as a percentage of total assets to capture the impact of investment on dividend payout. Companies in the growth stage of their life cycle tend to invest more in R&D and expand the firm size. As a result, high investment and capital expenditure decrease the probability of dividend payout (Fama&French, 2001).

To control for company growth rate, we use the logarithm of the total assets of the company. Companies with high growth rates and large company size are anticipated to pay dividends as, usually, they are mature companies with established constant earnings (Fenn&Liang, 2001). Another firm characteristic that needs to be considered is the proportion of total debt of the company used in the capital structure. We use total debt to total assets ratio of the company to control for leverage. High leverage leads to high sensitivity of firm performance to the business cycle. Moreover, high debt means high-interest expense and high financing costs which lead to low retained earnings available for dividend distribution. Descriptive statistics of control variables are presented in Table 1.

## 5. Empirical Results

The empirical results of the regression analysis for the three models outlined in Section 3 are presented in Table 2: OLS regression (column 1), OLS regression with year and company fixed effects (column 2) and OLS regression with year and industry fixed effects (column 3). Starting with the impact of the percentage of shares held by the CEO on the annual percentage change in dividend distributed, the relationship is found to be positive for all estimation models. The direction of the impact is economically intuitive and depicts a personal gain of the CEO from dividend distribution; yet, the coefficient is not statistically significant.

The impact of shares on dividend payout changes in times of recession. This negative relation is depicted by all three estimation models; however, coefficients are statistically insignificant. Although the negative impact may be counter-intuitive, the reason behind this change could be the following. The recession period is associated with a decrease in earnings and free cash flow to the firm. To protect the value of shares from further deterioration, the CEO may decide to retain excess cash for feasible losses and expenditures associated with recession rather than distribute it as a dividend.

In contrast to stock shares, stock options were found to have a very prominent impact on dividend payout. The result of OLS regression with year and industry fixed effects (column 3) indicates that a percentage increase in the number of options held by CEO will lead to a 2.8% of decrease in the annual dividends paid by the company. The coefficient is both economically and statistically significant and supports the hypothesis that the existence of stock options in managerial compensation portfolio decreases their dividend distribution incentives.

**Table 2: Regression results**

| DividendChange | (1) OLS | (2) OLS_FE_1 | (3) OLS_FE_2 |
|---|---|---|---|
| Shares | 0.758 | 0.414 | 0.713 |
|  | (0.473) | (0.484) | (0.485) |
| Options | -3.021** | -1.372 | -2.839** |
|  | (1.490) | (1.344) | (1.436) |
| RecShares | -0.746 | -0.821 | -0.688 |
|  | (0.516) | (0.656) | (0.522) |
| RecOptions | -2.411 | -2.455 | -2.563 |
|  | (3.408) | (5.325) | (3.597) |
| TCompensation | 5.682 | 14.513 | 4.743 |
|  | (5.922) | (8.840) | (6.011) |
| Recession | -7.847** | 1.898 | -12.128*** |
|  | (3.313) | (6.528) | (1.959) |
| Growth | -2.126 | -2.156 | -1.695 |
|  | (2.735) | (6.715) | (2.656) |
| Investment | -0.296 | -0.759 | -0.309 |
|  | (0.243) | (0.581) | (0.346) |
| Leverage | -17.370** | -35.832*** | -19.182*** |
|  | (6.786) | (9.508) | (6.816) |
| Profitability | 0.685*** | 0.869*** | 0.642*** |
|  | (0.124) | (0.258) | (0.126) |
| _cons | -0.197 | -75.244 | -19.689 |
|  | (15.922) | (81.177) | (16.728) |
| Obs. | 4600 | 4600 | 4600 |
| Pseudo $R^2$ | .z | .z | .z |

Standard errors are in parenthesis
*** p<0.01, ** p<0.05, * p<0.1

The underlying economic interpretation behind this relationship is the fact that options are not "dividend protected" which leads to deterioration of the option value in the case of dividend payout. The direction of the impact found in this analysis is aligned with existing literature [Fenn&Liang (2001), Lambert et al (1989)].

The coefficient for the main variable of interest of this paper, the impact of stock options held by CEO to annual dividend change distributed by the company in times of economic recession, is negative and economically intuitive. The direction and magnitude of impact do not change significantly during the recession in compare with the total sample period. However, the coefficient is not statistically significant, which implies that no proof has been found to support the hypothesis that stock options have a significantly negative impact on dividends during a recession.

The statistical insignificance of the main coefficient of interest may be due to the sample used for the analysis. The current analysis considers S&P 1500 companies which include large-, medium- and small-cap companies. However, the voting power of the CEO does differ depending on firm size. The dividend policy of large-cap companies is more likely to be determined by a decision of the board of directors, rather than the CEO. The voting power of CEOs of large-cap companies is also diminished by the impact of large shareholders and market expectations. Moreover, CEOs of large-cap companies usually hold an insignificant portion of shares outstanding which may impact the significance of the coefficient. Considering the possible impact of market capitalization on the voting power of CEO and CEO ownership, we have analyzed the underlying question by dividing the sample into three categories: large-cap, medium-cap, and small-cap companies. The results are presented and discussed in the next sub-section.

The coeffects for two control variables, leverage, and profitability, have predicted signs and statistically and economically significant. This supports the agency-based theories of dividend payout (Fenn&Liang, 2001). For example, a percentage change in the profitability of the company decreases annual dividends by 64 basis points, whereas, a percentage increase in leverage decreases dividends by 19% percent. The coefficient for growth variable is found to negatively impact annual dividend change. Companies with high growth opportunities may require funding for investment to sustain the current growth rate of assets which can be realized at the cost of decreased dividend distribution [ Rozeff (1982), White (1996)]. However, the coefficients for both growth and investment variables are statistically insignificant.

**5.1. The role of company size in the relationship between managerial stock ownership and dividend distribution**

It has been determined that to decrease transaction costs, large and institutional investors prefer to invest in large-cap companies. As these investors prefer dividends, they create a "dividend clientele" effect for the company. Consequently, large companies choose to pay dividends, whereas, small companies do not (Redding, 1997). Consequently, in comparison with small-cap companies, in large-cap companies, institutional shareholders have more decisive power over dividend policy.

A decrease in the voting power of management in large companies may alter the significance of the impact of CEO stock ownership on the dividend payout policy of the company. In the time of recession, management may be required to continue paying dividends, although this may alter the value of managerial stock options and not align with their true incentives. To investigate the impact of company size on the relationship between managerial ownership and dividend policy, specifically during the recession period, we have divided the full sample into 3 sub-samples: S&P 400 medium-cap companies, S&P 600 small-cap companies, and S&P 500 large-cap companies. Descriptive statistics of the variables of each sub-sample are presented in the Appendix. we have used the same regression model as in the main analysis and performed three separate regressions by applying the OLS regression model with year and industry fixed effects.

**Table 3: Regression results**

|  | (1) | (2) | (3) |
|---|---|---|---|
| DividendChange | S&P600 | S&P400 | S&P500 |
| Shares | 0.995 | 0.070 | 0.464* |
|  | (0.959) | (0.140) | (0.255) |
| Options | 0.652 | -0.171 | -4.832* |
|  | (5.817) | (1.361) | (2.818) |
| RecShares | 2.110 | -0.137 | -0.812** |
|  | (2.890) | (0.162) | (0.362) |
| RecOptions | -4.096 | -25.637** | 4.836 |
|  | (10.625) | (11.263) | (7.257) |
| TCompensation | 23.651** | 1.818 | 8.676** |
|  | (10.980) | (3.304) | (4.360) |
| Recession | -25.743 | -3.556 | -11.834*** |
|  | (17.829) | (8.767) | (4.016) |
| Growth | -29.521 | 7.794 | -3.694 |
|  | (27.844) | (6.538) | (2.539) |
| Investment | 1.190 | -1.003** | -0.839* |
|  | (0.740) | (0.395) | (0.462) |
| Leverage | 59.568 | -15.557*** | -22.74*** |
|  | (72.625) | (4.739) | (7.211) |
| Profitability | 1.787*** | 1.009*** | 0.869*** |
|  | (0.629) | (0.344) | (0.252) |
| _cons | 297.977 | -74.970 | -3.128 |
|  | (211.257) | (62.962) | (26.326) |
| Obs. | 2114 | 2048 | 3549 |
| Pseudo $R^2$ | .z | .z | .z |

Standard errors are in parenthesis

*** $p<0.01$, ** $p<0.05$, * $p<0.1$

The results of regressions support the hypothesis of the impact of company size on managerial decision power over dividends (Table 3). Both the sign and significance of the coefficient of the main variable of interest, i.e. RecOptions, are found to be changing depending on company size. For small and medium-sized companies, the sign is negative as expected, whereas for large-sized companies it is found to be positive. This indicates to a possible decrease in the voting power of the CEO in large-cap companies. The coefficient is both economically and statistically significant for medium-sized companies which implies that a percentage increase in stock options held by the CEO during a recession will lead to 25% decrease in annual dividends paid.

The impact of CEO shares is especially pronounced for S&P 500 large-cap companies. The impact is positive and significant for the full sample reflecting the personal gains of the CEO from dividend distribution as an ordinary investor. However, the direction of the impact changes in times of the recession. One of the reasons causing such an impact is the willingness of the CEO to maintain enough cash and liquidity in the company. Varying directions of the impact of the shares and options held by CEO for small-, medium- and large-cap companies indicate to the existing influence of company size on the relationship between managerial stock ownership and dividend payout decisions.

## 6. Conclusion

This paper studies how the business cycle impacts the relationship between managerial stock incentives and dividend distribution of companies by using the data of S&P 1500 non-financial companies during 2000-2018. The findings of the OLS regression model with year and industry fixed effects support existing literature by determining the negative impact of stock options held by the CEO on dividend payout. However, no evidence has been found to support a significant relationship between variables during a recession period. We also find evidence that company size influences the role of the business cycle in the relationship between managerial stock incentives and dividends distribution. The stock options held by CEOs of medium-cap companies during the recession have a negative and significant impact on dividends, whereas, this impact is positive for large-cap companies. This supports the hypothesis that the business cycle impacts the relationship between managerial stock incentives and dividend distribution, and this impact largely depends on the size of the company.

The results imply that medium-cap companies experience high agency cost problems, as composition of CEO compensation alter dividend policy of the company during economic contraction. The finding is specifically important for dividend-oriented investors in their design of the applicable investment portfolio to match their target dividend gain.

# 7. Appendix

## 7.1. Additional Summary Statistics

Descriptive statistics of the S&P 600 Small-cap, S&P 400 Mid-cap and S&P 500 Large-cap companies are presented in Table 4. On average CEOs of small-cap companies hold the largest number of shares and options as a percentage of shares outstanding in compare with large and

## Table 4: Descriptive Statistics

**S&P 600 Small Cap Companies**

| Variable | Obs | Mean | Std.Dev. | Min | Max |
|---|---|---|---|---|---|
| Shares | 5687 | 3.615 | 8.503 | 0 | 77.56 |
| Options | 5747 | .166 | .412 | 0 | 13.444 |
| Growth | 7155 | 8.731 | .514 | 5.936 | 10.421 |
| Investment | 7140 | 5.251 | 7.107 | 0 | 298.956 |
| Leverage | 6985 | .447 | .963 | .001 | 74.934 |
| Profitab | 7148 | 7.628 | 95.127 | -7956.845 | 92.041 |

**S&P 400 Mid Cap Companies**

| Variable | Obs | Mean | Std.Dev. | Min | Max |
|---|---|---|---|---|---|
| Shares | 3996 | 2.415 | 6.613 | 0 | 67.386 |
| Options | 4055 | .165 | .464 | 0 | 18.206 |
| Growth | 4608 | 9.232 | .483 | 7.174 | 10.629 |
| Investment | 4589 | 5.551 | 6.65 | -3.27 | 81.527 |
| Leverage | 4437 | .516 | .317 | .029 | 4.35 |
| Profitab | 4604 | 11.304 | 8.279 | -93.745 | 60.991 |

**S&P 500 Large Cap Companies**

| Variable | Obs | Mean | Std.Dev. | Min | Max |
|---|---|---|---|---|---|
| Shares | 5561 | 1.321 | 4.7 | 0 | 75.501 |
| Options | 5628 | .105 | .3 | 0 | 10.171 |
| Growth | 6055 | 9.908 | .626 | 7.586 | 11.902 |
| Investment | 6033 | 4.887 | 4.737 | -.062 | 60.552 |
| Leverage | 5854 | .529 | .221 | .011 | 2.917 |
| Profitab | 6052 | 12.232 | 8.306 | -83.332 | 65.942 |

mid-cap companies. This is largely because small-cap companies have less shares outstanding and a smaller number of large shareholders. In contrast, on average CEOs of large-cap companies hold the least number of shares and options. One of the reasons causing this is the willingness of large shareholders to decrease insider ownership and limit the voting power of the CEO.